\documentstyle[12pt,a4,axodraw]{article}

\newcommand\order{\mathop{\cal O}\nolimits}
\newcommand\dPS{d\,P\!S}
\newcommand\setscale[1]{\def\axoscale{#1 }}
\newcommand{\lsim}{\stackrel{\lower.7ex\hbox{$<$}}{\lower.7ex\hbox{$\sim$}}}
\newcommand\GeV{\mbox{ GeV}}
\newcommand\GEV{\mbox{GeV}}

\newcommand\PB{\mbox{pb}}
\begin{document}


\thispagestyle{empty}
\setcounter{page}{0}

\begin{flushright}
PSI-PR-94-06\\
February 1994
\end{flushright}

\vspace*{\fill}

\begin{center}
{\Large\bf WWF: a generator for $e^+e^- \to 4\;\mbox{fermions} + \gamma$}\\
\vspace{2em}
\large
\begin{tabular}[t]{c}
Geert~Jan van~Oldenborgh\\
Paula~J.~Franzini\\
Arianna~Borrelli\\
\\
{\it Paul Scherrer Institut, CH-5232 Villigen PSI, Switzerland}\\
\end{tabular}
\end{center}

\vspace*{\fill}

\begin{abstract}\noindent
We present an efficient generator for the process $e^+e^- \to
4\;\mbox{fermions} + \gamma$ through off-shell $W$ pairs.
It is based on a massless matrix element with leading $\order(m^2)$
corrections.  Only the resonant $WW$ graphs are included.  We have tested it
against a matrix element without these approximations and found agreement to
within $\sim$1\% at LEPII energies.
\end{abstract}

\vspace*{\fill}
{\noindent\footnotesize Email addresses: gj@csun.psi.ch,
franzini@psiclu.cern.ch and borrelli@psiclu.cern.ch.}
\newpage


\section{Introduction}

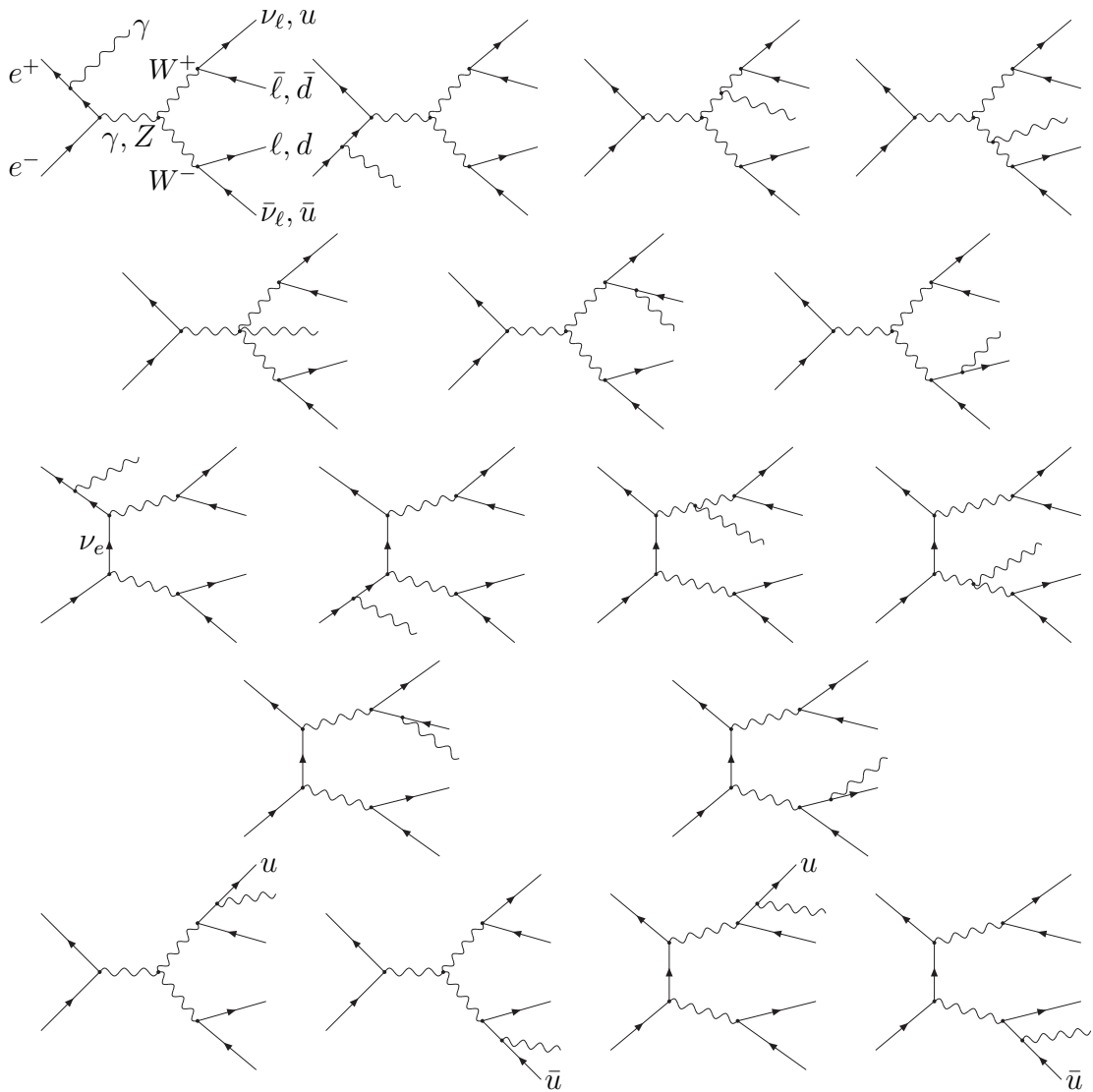
\begin{figure}
\setscale{.75}
\unitlength .75bp
\centerline{
\begin{picture}(120,100)(0,0)
\ArrowLine(0,20)(30,50)
\put(0,20){\makebox(0,0)[br]{$e^-$}}
\ArrowLine(30,50)(15,65)
\Vertex(15,65){1}
\Photon(15,65)(45,95){2}{4}
\put(47,95){\makebox(0,0)[l]{$\gamma$}}
\ArrowLine(15,65)(0,80)
\put(0,80){\makebox(0,0)[tr]{$e^+$}}
\Vertex(30,50){1}
\Photon(30,50)(60,50){2}{3}
\put(45,46){\makebox(0,0)[t]{$\gamma,Z$}}
\Vertex(60,50){1}
\Photon(60,50)(80,75){2}{4}
\put(80,70){\makebox(0,0)[br]{\small$W^+$}}
\Vertex(80,75){1}
\ArrowLine(115,65)(80,75)
\put(117,65){\makebox(0,0)[l]{$\bar{\ell},\bar{d}$}}
\ArrowLine(80,75)(110,100)
\put(112,100){\makebox(0,0)[l]{$\nu_{\ell},u$}}
\Photon(60,50)(80,25){2}{4}
\put(80,25){\makebox(0,0)[tr]{\small$W^-$}}
\Vertex(80,25){1}
\ArrowLine(110,0)(80,25)
\put(112,0){\makebox(0,0)[l]{$\bar{\nu}_{\ell},\bar{u}$}}
\ArrowLine(80,25)(115,35)
\put(117,35){\makebox(0,0)[l]{$\ell,d$}}
\end{picture}
\hfill
\begin{picture}(120,100)(0,0)
\ArrowLine(0,20)(15,35)
\ArrowLine(15,35)(30,50)
\Vertex(15,35){1}
\Photon(15,35)(45,15){2}{4}
\ArrowLine(30,50)(0,80)
\Vertex(30,50){1}
\Photon(30,50)(60,50){2}{3}
\Vertex(60,50){1}
\Photon(60,50)(80,75){2}{4}
\Vertex(80,75){1}
\ArrowLine(115,65)(80,75)
\ArrowLine(80,75)(110,100)
\Photon(60,50)(80,25){2}{4}
\Vertex(80,25){1}
\ArrowLine(110,0)(80,25)
\ArrowLine(80,25)(115,35)
\end{picture}
\hfill
\begin{picture}(120,100)(0,0)
\ArrowLine(0,20)(30,50)
\ArrowLine(30,50)(0,80)
\Vertex(30,50){1}
\Photon(30,50)(60,50){2}{3}
\Vertex(60,50){1}
\Photon(60,50)(80,75){2}{4}
\Vertex(70,62.5){1}
\Photon(70,62.5)(108,50){2}{4}
\Vertex(80,75){1}
\ArrowLine(115,65)(80,75)
\ArrowLine(80,75)(110,100)
\Photon(60,50)(80,25){2}{4}
\Vertex(80,25){1}
\ArrowLine(110,0)(80,25)
\ArrowLine(80,25)(115,35)
\end{picture}
\hfill
\begin{picture}(120,100)(0,0)
\ArrowLine(0,20)(30,50)
\ArrowLine(30,50)(0,80)
\Vertex(30,50){1}
\Photon(30,50)(60,50){2}{3}
\Vertex(60,50){1}
\Photon(60,50)(80,75){2}{4}
\Vertex(80,75){1}
\ArrowLine(115,65)(80,75)
\ArrowLine(80,75)(110,100)
\Photon(60,50)(80,25){2}{4}
\Vertex(70,37.5){1}
\Photon(70,37.5)(108,50){2}{4}
\Vertex(80,25){1}
\ArrowLine(110,0)(80,25)
\ArrowLine(80,25)(115,35)
\end{picture}
} 
\medskip
\centerline{
\hfill
\begin{picture}(120,100)(0,0)
\ArrowLine(0,20)(30,50)
\ArrowLine(30,50)(0,80)
\Vertex(30,50){1}
\Photon(30,50)(60,50){2}{3}
\Vertex(60,50){1}
\Photon(60,50)(100,50){2}{4}
\Photon(60,50)(80,75){2}{4}
\Vertex(80,75){1}
\ArrowLine(115,65)(80,75)
\ArrowLine(80,75)(110,100)
\Photon(60,50)(80,25){2}{4}
\Vertex(80,25){1}
\ArrowLine(110,0)(80,25)
\ArrowLine(80,25)(115,35)
\end{picture}
\hfill
\begin{picture}(120,100)(0,0)
\ArrowLine(0,20)(30,50)
\ArrowLine(30,50)(0,80)
\Vertex(30,50){1}
\Photon(30,50)(60,50){2}{3}
\Vertex(60,50){1}
\Photon(60,50)(80,75){2}{4}
\Vertex(80,75){1}
\ArrowLine(120,65)(96,71)
\Line(96,71)(80,75)
\Vertex(96,71){1}
\Photon(96,71)(115,50){-2}{3}
\ArrowLine(80,75)(110,100)
\Photon(60,50)(80,25){2}{4}
\Vertex(80,25){1}
\ArrowLine(110,0)(80,25)
\ArrowLine(80,25)(115,35)
\end{picture}
\hfill
\begin{picture}(120,100)(0,0)
\ArrowLine(0,20)(30,50)
\ArrowLine(30,50)(0,80)
\Vertex(30,50){1}
\Photon(30,50)(60,50){2}{3}
\Vertex(60,50){1}
\Photon(60,50)(80,75){2}{4}
\Vertex(80,75){1}
\ArrowLine(115,65)(80,75)
\ArrowLine(80,75)(110,100)
\Photon(60,50)(80,25){2}{4}
\Vertex(80,25){1}
\ArrowLine(110,0)(80,25)
\Line(80,25)(96,29)
\ArrowLine(96,29)(120,35)
\Vertex(96,29){1}
\Photon(96,29)(115,50){2}{3}
\end{picture}
\hfill
} 
\medskip
\centerline{
\begin{picture}(110,100)(10,0)
\ArrowLine(10,10)(45,35)
\ArrowLine(45,35)(45,65)
\put(43,50){\makebox(0,0)[r]{$\nu_e$}}
\ArrowLine(45,65)(27.5,77.5)
\ArrowLine(27.5,77.5)(10,90)
\Vertex(27.5,77.5){1}
\Photon(27.5,77.5)(60,95){2}{4}
\Vertex(45,35){1}
\Vertex(45,65){1}
\Photon(45,65)(80,75){2}{4}
\Vertex(80,75){1}
\ArrowLine(115,65)(80,75)
\ArrowLine(80,75)(110,100)
\Photon(45,35)(80,25){2}{4}
\Vertex(80,25){1}
\ArrowLine(110,0)(80,25)
\ArrowLine(80,25)(115,35)
\end{picture}
\hfill
\begin{picture}(110,100)(10,0)
\Vertex(27.5,22.5){1}
\Photon(27.5,22.5)(60,5){2}{4}
\ArrowLine(10,10)(27.5,22.5)
\ArrowLine(27.5,22.5)(45,35)
\ArrowLine(45,35)(45,65)
\ArrowLine(45,65)(10,90)
\Vertex(45,35){1}
\Vertex(45,65){1}
\Photon(45,65)(80,75){2}{4}
\Vertex(80,75){1}
\ArrowLine(115,65)(80,75)
\ArrowLine(80,75)(110,100)
\Photon(45,35)(80,25){2}{4}
\Vertex(80,25){1}
\ArrowLine(110,0)(80,25)
\ArrowLine(80,25)(115,35)
\end{picture}
\hfill
\begin{picture}(110,100)(10,0)
\ArrowLine(10,10)(40,35)
\ArrowLine(40,35)(40,65)
\ArrowLine(40,65)(10,90)
\Vertex(40,35){1}
\Vertex(40,65){1}
\Photon(40,65)(80,75){2}{5}
\Vertex(60,70){1}
\Photon(60,70)(95,50){-2}{5}
\Vertex(80,75){1}
\ArrowLine(115,65)(80,75)
\ArrowLine(80,75)(110,100)
\Photon(40,35)(80,25){2}{5}
\Vertex(80,25){1}
\ArrowLine(110,0)(80,25)
\ArrowLine(80,25)(115,35)
\end{picture}
\hfill
\begin{picture}(110,100)(10,0)
\ArrowLine(10,10)(40,35)
\ArrowLine(40,35)(40,65)
\ArrowLine(40,65)(10,90)
\Vertex(40,35){1}
\Vertex(40,65){1}
\Photon(40,65)(80,75){2}{5}
\Vertex(80,75){1}
\ArrowLine(115,65)(80,75)
\ArrowLine(80,75)(110,100)
\Photon(40,35)(80,25){2}{5}
\Vertex(60,30){1}
\Photon(60,30)(95,50){-2}{5}
\Vertex(80,25){1}
\ArrowLine(110,0)(80,25)
\ArrowLine(80,25)(115,35)
\end{picture}
} 
\medskip
\centerline{
\hfill
\begin{picture}(110,100)(10,0)
\ArrowLine(10,10)(40,35)
\ArrowLine(40,35)(40,65)
\ArrowLine(40,65)(10,90)
\Vertex(40,35){1}
\Vertex(40,65){1}
\Photon(40,65)(75,75){2}{4}
\Vertex(75,75){1}
\ArrowLine(115,65)(91,71)
\Line(91,71)(75,75)
\Vertex(91,71){1}
\Photon(91,71)(120,50){-2}{3.5}
\ArrowLine(75,75)(110,100)
\Photon(40,35)(75,25){2}{4}
\Vertex(75,25){1}
\ArrowLine(110,0)(75,25)
\ArrowLine(75,25)(115,35)
\end{picture}
\hfill
\begin{picture}(110,100)(10,0)
\ArrowLine(10,10)(40,35)
\ArrowLine(40,35)(40,65)
\ArrowLine(40,65)(10,90)
\Vertex(40,35){1}
\Vertex(40,65){1}
\Photon(40,65)(75,75){2}{4}
\Vertex(75,75){1}
\ArrowLine(115,65)(75,75)
\ArrowLine(75,75)(110,100)
\Photon(40,35)(75,25){2}{4}
\Vertex(75,25){1}
\ArrowLine(110,0)(75,25)
\Line(75,25)(91,29)
\ArrowLine(91,29)(115,35)
\Vertex(91,29){1}
\Photon(91,29)(120,50){2}{3.5}
\end{picture}
\hfill
} 
\medskip
\centerline{
\begin{picture}(120,100)(0,0)
\ArrowLine(0,20)(30,50)
\ArrowLine(30,50)(0,80)
\Vertex(30,50){1}
\Photon(30,50)(60,50){2}{3}
\Vertex(60,50){1}
\Photon(60,50)(80,75){2}{4}
\Vertex(80,75){1}
\ArrowLine(115,65)(80,75)
\Line(80,75)(90,85)
\ArrowLine(90,85)(110,105)
\put(112,105){\makebox(0,0)[l]{$u$}}
\Vertex(90,85){1}
\Photon(90,85)(120,90){-2}{3.5}
\Photon(60,50)(80,25){2}{4}
\Vertex(80,25){1}
\ArrowLine(110,0)(80,25)
\ArrowLine(80,25)(115,35)
\end{picture}
\hfill
\begin{picture}(120,100)(0,0)
\ArrowLine(0,20)(30,50)
\ArrowLine(30,50)(0,80)
\Vertex(30,50){1}
\Photon(30,50)(60,50){2}{3}
\Vertex(60,50){1}
\Photon(60,50)(80,75){2}{4}
\Vertex(80,75){1}
\ArrowLine(115,65)(80,75)
\ArrowLine(80,75)(110,100)
\Photon(60,50)(80,25){2}{4}
\Vertex(80,25){1}
\ArrowLine(110,-5)(90,15)
\put(112,-5){\makebox(0,0)[l]{$\bar{u}$}}
\Line(90,15)(80,25)
\ArrowLine(80,25)(115,35)
\Vertex(90,15){1}
\Photon(90,15)(120,10){2}{3.5}
\end{picture}
\hfill
\begin{picture}(110,100)(10,0)
\ArrowLine(10,10)(40,35)
\ArrowLine(40,35)(40,65)
\ArrowLine(40,65)(10,90)
\Vertex(40,35){1}
\Vertex(40,65){1}
\Photon(40,65)(75,75){2}{4}
\Vertex(75,75){1}
\ArrowLine(115,65)(75,75)
\Line(75,75)(85,85)
\ArrowLine(85,85)(105,105)
\put(107,105){\makebox(0,0)[l]{$u$}}
\Vertex(85,85){1}
\Photon(85,85)(120,80){-2}{4}
\Photon(40,35)(75,25){2}{4}
\Vertex(75,25){1}
\ArrowLine(110,0)(75,25)
\ArrowLine(75,25)(115,35)
\end{picture}
\hfill
\begin{picture}(110,100)(10,0)
\ArrowLine(10,10)(40,35)
\ArrowLine(40,35)(40,65)
\ArrowLine(40,65)(10,90)
\Vertex(40,35){1}
\Vertex(40,65){1}
\Photon(40,65)(75,75){2}{4}
\Vertex(75,75){1}
\ArrowLine(115,65)(75,75)
\ArrowLine(75,75)(110,100)
\Photon(40,35)(75,25){2}{4}
\Vertex(75,25){1}
\ArrowLine(105,-5)(85,15)
\put(107,-5){\makebox(0,0)[l]{$\bar{u}$}}
\Line(85,15)(75,25)
\Vertex(85,15){1}
\Photon(85,15)(120,20){2}{4}
\ArrowLine(75,25)(115,35)
\end{picture}
} 
\caption{The Feynman diagrams included in the generator.  (The last row only
occurs when the $W$ decays hadronically).}
\label{fig:feynman}
\end{figure}

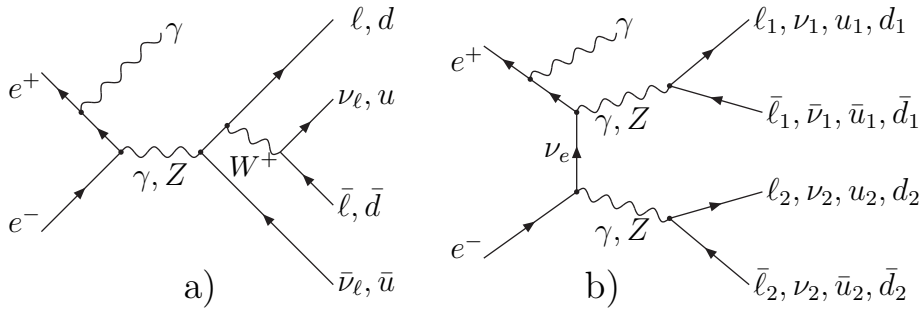
\begin{figure}[htb]
\centerline{
\hfill
\begin{picture}(120,100)(0,0)
\ArrowLine(0,20)(30,50)
\put(0,20){\makebox(0,0)[br]{$e^-$}}
\ArrowLine(30,50)(15,65)
\Vertex(15,65){1}
\Photon(15,65)(45,95){2}{4}
\put(47,95){\makebox(0,0)[l]{$\gamma$}}
\ArrowLine(15,65)(0,80)
\put(0,80){\makebox(0,0)[tr]{$e^+$}}
\Vertex(30,50){1}
\Photon(30,50)(60,50){2}{3}
\put(45,46){\makebox(0,0)[t]{$\gamma,Z$}}
\Vertex(60,50){1}
\ArrowLine(110,0)(60,50)
\put(112,0){\makebox(0,0)[l]{$\bar{\nu}_{\ell},\bar{u}$}}
\Line(60,50)(70,60)
\Vertex(70,60){1}
\Photon(70,60)(90,50){-2}{2.5}
\put(90,50){\makebox(0,0)[tr]{\small$W^+$}}
\ArrowLine(110,30)(90,50)
\put(112,30){\makebox(0,0)[l]{$\bar{\ell},\bar{d}$}}
\ArrowLine(90,50)(110,70)
\put(112,70){\makebox(0,0)[l]{$\nu_{\ell},u$}}
\ArrowLine(70,60)(110,100)
\put(117,100){\makebox(0,0)[l]{$\ell,d$}}
\put(60,-5){\makebox(0,0)[b]{\large a)}}
\end{picture}
\hfill
\begin{picture}(150,100)(10,0)
\ArrowLine(10,10)(45,35)
\put(10,10){\makebox(0,0)[br]{$e^-$}}
\ArrowLine(45,35)(45,65)
\put(43,50){\makebox(0,0)[r]{$\nu_e$}}
\ArrowLine(45,65)(27.5,77.5)
\ArrowLine(27.5,77.5)(10,90)
\put(10,90){\makebox(0,0)[tr]{$e^+$}}
\Vertex(27.5,77.5){1}
\Photon(27.5,77.5)(60,95){2}{4}
\put(60,97){\makebox(0,0)[l]{$\gamma$}}
\Vertex(45,35){1}
\Vertex(45,65){1}
\Photon(45,65)(80,75){2}{4}
\put(62.5,67){\makebox(0,0)[t]{$\gamma,Z$}}
\Vertex(80,75){1}
\ArrowLine(115,65)(80,75)
\put(117,65){\makebox(0,0)[l]{$\bar{\ell}_1,\bar{\nu}_1,\bar{u}_1,\bar{d}_1$}}
\ArrowLine(80,75)(110,100)
\put(112,100){\makebox(0,0)[l]{$\ell_1,\nu_1,u_1,d_1$}}
\Photon(45,35)(80,25){2}{4}
\put(62.5,25){\makebox(0,0)[t]{$\gamma,Z$}}
\Vertex(80,25){1}
\ArrowLine(110,0)(80,25)
\put(112,0){\makebox(0,0)[l]{$\bar{\ell}_2,\nu_2,\bar{u}_2,\bar{d}_2$}}
\ArrowLine(80,25)(115,35)
\put(117,35){\makebox(0,0)[l]{$\ell_2,\nu_2,u_2,d_2$}}
\put(55,-5){\makebox(0,0)[b]{\large b)}}
\end{picture}
\hfill
}
\medskip
\caption[]{Some diagrams which are {\em not\/} included.  Diagram a) is a
non-resonant diagram, which is present in all channels but suppressed by a
factor $\Gamma_W/m_W$.  Diagram b) occurs only in certain channels; it peaks in
different regions of phase space than the $W$ graphs.}
\label{fig:notincluded}
\end{figure}

One of the main motivations of the LEPII electron-positron collider is the
study of pair-produced $W$ bosons.   In order for the anticipated
high precision ($\sim$ 1\%) of the measurements to be meaningful, they must be
complemented by a complete calculation of the electroweak radiative
corrections to one-loop level, which are of a similar order of magnitude. A
subset of these are the hard radiative processes, in which a detectable photon
is emitted.  This process has been studied before for the production and decay
of on-shell $W$ bosons (for production see, e.g.,
\cite{Jochem&Fred&KarolW,Grace}).  However, the width of the $W$ boson is such
that one has to consider off-shell effects by looking at the complete process
$e^+e^-\to4\;\mbox{fermions} + \gamma$.  For most of phase space this is still
dominated by the resonant $W$ diagrams shown in Fig.~\ref{fig:feynman}.  The
non-resonant $W$ diagrams (an example is shown in Fig.~\ref{fig:notincluded}a)
are suppressed by factors $\Gamma_W/m_W$ and can safely be neglected.  The
$\gamma\gamma$, $\gamma Z$ and $ZZ$ graphs shown in
Fig.~\ref{fig:notincluded}b occur only when particle--anti-particle pairs are
present in the final state.  They also peak in a different region of phase
space and can thus be excluded by cuts on resonant $Z$'s and $\gamma$'s
decaying into observable particles.

The non-collinear behaviour of the resonant graphs of Fig.~\ref{fig:feynman}
was already given in Ref.~\cite{Andre&DanielBrems}.  We have extended this
calculation to the full phase space (including mass effects \cite{KleissMass})
and all decay channels, added initial state collinear bremsstrahlung
\cite{LeidenEemumu2loop} and converted it into an efficient event generator
with an interface to Jetset \cite{jetset73}. We also verified that the matrix
element agrees with an independent approach \cite{Grace}, and checked that the
contribution of the other graphs is indeed small.

\section{Method}

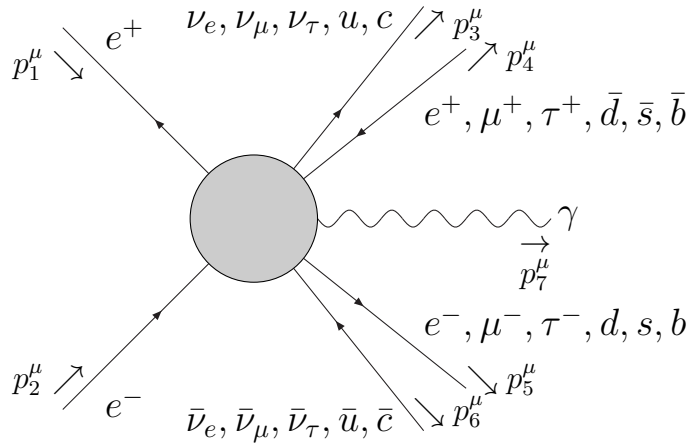
\begin{figure}[htb]
\begin{center}
\unitlength .8bp
\setscale{.8}
\begin{picture}(230,200)(0,0)
\ArrowLine( 10, 10)(100,100)
\put( 20, 20){\makebox(0,0)[br]{$p_2^\mu\nearrow$}}
\put( 30, 20){\makebox(0,0)[tl]{\large $e^-$}}
\ArrowLine(100,100)( 10,190)
\put( 20,180){\makebox(0,0)[tr]{$p_1^\mu\searrow$}}
\put( 30,180){\makebox(0,0)[bl]{\large $e^+$}}
\ArrowLine(180,  0)(100,100)
\put(175,  5){\makebox(0,0)[bl]{$\searrow p_6^\mu$}}
\put(165, 10){\makebox(0,0)[tr]{\large $\bar{\nu}_e,\bar{\nu}_\mu,
    \bar{\nu}_\tau,\bar{u},\bar{c}$}}
\ArrowLine(100,100)(200, 20)
\put(200, 20){\makebox(0,0)[bl]{$\searrow p_5^\mu$}}
\put(180, 45){\makebox(0,0)[bl]{\large $e^-,\mu^-,\tau^-,d,s,b$}}
\ArrowLine(200,180)(100,100)
\put(200,184){\makebox(0,0)[tl]{$\nearrow p_4^\mu$}}
\put(180,159){\makebox(0,0)[tl]{\large $e^+,\mu^+,\tau^+,\bar{d},\bar{s},
    \bar{b}$}}
\ArrowLine(100,100)(180,200)
\put(175,199){\makebox(0,0)[tl]{$\nearrow p_3^\mu$}}
\put(165,190){\makebox(0,0)[br]{\large $\nu_e,\nu_\mu,\nu_\tau,u,c$}}
\Photon(100,100)(240,100){4}{7}
\put(240, 90){\makebox(0,0)[tr]{\shortstack{$\rightarrow$\\$p_7^\mu$}}}
\put(243,100){\makebox(0,0)[l]{\large $\gamma$}}
\GCirc(100,100){30}{.8}
\end{picture}
\end{center}
\caption{Assignment of the momenta}
\label{fig:diagram}
\end{figure}

In this section we will sketch the way the program is built up and which
methods are used to compute the various ingredients which make up the
generator.  This consists of the event generator package, a front end that
chooses the final state particles, the phase space routines that transform
the random variables to four-vectors and the matrix element, including mass
effects and QCD corrections.  Each of these parts is discussed in turn.
Throughout we will use the notation of Fig.\ \ref{fig:diagram}.

\subsection{Event generator}

We use two approaches for the generation of weight one events which are useful
for a full physics study.\footnote{There is also the possibility of using
weighted events (normal Monte Carlo integration) for studies for which
time-consuming tasks such as jet fragmentation and detector simulation are not
needed.}  The first possibility is to throw random points into the integration
region and accept an event if the ratio of its weight to the maximal weight
exceeds yet another random number in $(0,1)$.  This depends on a knowledge of
the actual maximum weight, for which we take 20\% more than the largest weight
found in the first 500 events, which are discarded.  As the integrand has been
smoothed out by mappings  so that the function does not fluctuate much this
estimate normally exceeds the maximum weight.  If a larger weight is found
during the run the event is just accepted once for sure, and a second time
with a probability depending on the ratio of the excess to the maximum weight.
 This approach has been implemented in the routine {\tt simplemc} called by
{\tt simple}.  It has an efficiency (events generated/events accepted) of
about 5\%.

A more sophisticated method is to use the grid produced by an adaptive Monte
Carlo algorithm such as Vegas \cite{LepageVEGAS} or Bases \cite{bases}.  These
adaptive algorithms use a discrete mapping on each integration variable to
reduce the variance of the integrand; generating points using the same
discrete mapping improves the efficiency of the event generator.  Event
generation thus proceeds in two steps: first one finds a good grid by running
the integration program, next one generates the events.  The maximal weight
has also been found already by the integration step.  We integrated our
programs with the Spring event generator, which builds on the Bases
integration routine.  The efficiency of this event generator can be as high as
20\%, but this depends very much on the amount of time used to generate the
grid.

\subsection{Front end}

The function which is called by the integration or event generation package is
{\tt wwf}.  This takes as argument a point in a 13-dimensional unit hyper-cube
and returns the weight associated with that point in picobarn.  The
corresponding four-vectors and particles are stored in common as described in
section \ref{subsec:jetset} for further treatment in the routine {\tt spevnt}
(jet fragmentation, detector simulation, etc.).

The first activity is the conversion of the user-supplied values for the
physics constants to coupling constants, $W$ width and cut-off angles.  This
is repeated whenever these values change, printing the new values.  Next the
first semi-random variable is used to select the decay channel --- leptonic,
semi-leptonic or hadronic --- from among the allowed channels specified by the
user in {\tt wwfset} (see section \ref{subsec:parameters}).  After rescaling
the variable, it is reused to select the leptonic decay channels ($e,\mu,\tau$;
again from those allowed by the user) and the hadronic channels (weighted
according to the Cabibbo-Kobayashi-Maskawa (CKM) matrix elements).  One more
choice needs to be made: the multi-channel integration algorithm means that
the cross section is a sum of 4 (leptonic) to 6 (hadronic) different channels
(in each channel the program focuses on the radiation off one of
the external legs, see Ref.\ \cite{Andre&DanielBrems}).  The same first random
variable is again rescaled and used to choose the channel; finally the
corresponding routine {\tt wg1a}-{\tt wg2b} is called with 11 random variables.

In the case that exponentiated collinear initial state bremsstrahlung is added
to the full one-photon calculation the routine {\tt wwfini} computes the
relevant structure function factor \cite{LeidenEemumu2loop} and generates the
initial state momenta after this radiation using the remaining random variable.
 All the momentum is assumed to be lost by just one of the incoming particles,
although the function used includes up to two hard photons.  This will only
make a difference of order $\alpha^2$ (with respect to the radiative cross
section) when the cuts used discriminate between differently boosted events.

\subsection{Phase space}

There are thus 6 different weight-computing routines, which each call their
corresponding phase space routine {\tt five1a}-{\tt five2b}.  This phase space
routine is called twice to construct a massive ($p_i^\mu$) and massless
($q_i^\mu$) set of variables.  The latter is used to evaluate the matrix
element consistently, the former to cancel the collinear poles and to
implement the experimental cuts.  For the denominator, {\tt wg1a}-{\tt wg2b}
compute the jacobians of the other channels from the four-vectors with the
routines {\tt g1a}-{\tt g2b}.  When the event satisfies some simple
kinematical cuts in {\tt wwfrej} the massless matrix element {\tt mat} is
computed.  This is the same for all channels, but different mass corrections
{\tt getmm1}-{\tt getmm2} are added. Finally histograms are filled ({\tt
wwfill}).

The phase space routines evaluate the phase space element as follows.
\begin{description}
\item[{\tt five1a,g1a}:] radiation off particle 1 (the incoming positron).
The phase space integral is evaluated as
\begin{equation}
    \dPS_{1a} = \frac{1}{(2\pi)^3} \frac{E_7\,dE_7}{2} \; d\phi_7 \;
        d\cos\theta_7 \; \dPS'
\end{equation}
\begin{equation}
    \dPS' = \frac{1}{2\pi}ds^+\; \frac{1}{2\pi}ds^- \; \dPS_2(s,s^+,s^-)
        \; \dPS_2(s^+,m_3^2,m_4^2) \; \dPS_2(s^-,m_5^2,m_6^2) \;,
\end{equation}
where $E_i$, $\theta_i$ and $\phi_i$ are the energy and angles of particle $i$
in the lab frame.
In this equation $s^\pm$ are the momenta squared of the off-shell $W^\pm$ and
\begin{equation}
    \dPS_2(s,s^+,s^-) =
        d\phi^W d\cos\theta^W \frac{\sqrt{\lambda(s,s^+,s^-)}}{32\pi^2s}\;,
\end{equation}
with these angles defined in the $W^+W^-$ CM frame.  The K\"all\'en function
$\lambda$ is defined as $\lambda(x,y,z) = (x-y-z)^2-4yz$.  Analogous
expressions hold for the other two 2-particle phase space elements.  We map
away the $1/t$, $1/E_7$ and $1/(E_1-|\vec{p_1}|\cos\theta_7) \propto 1/(p_1
p_7)$ behaviours of the matrix element, taking into account the cuts on the
photon energy and angle in the first two.  The mappings are defined in {\tt
five1a} and {\tt mapt}.

\item[{\tt five1b,g1b}:] radiation off particle 2 (the incoming electron).
These routines are just the mirror image of the previous ones.

\item[{\tt five2a(0),g2a(0)}:] radiation off particle 4 (the outgoing $e^+,
\mu^+,\tau^+,\bar{d},\bar{s}$ or $\bar{b}$).  The phase space integral is now
evaluated as
\begin{equation}
    \dPS_{2a} = \frac{1}{2\pi}ds^+\; \frac{1}{2\pi}ds^- \; \dPS_2(s,s^+,s^-)
        \; \dPS_3(s^+,m_3^2,m_4^2) \; \dPS_2(s^-,m_5^2,m_6^2) \;,
\end{equation}
with the 3-particle Dalitz decay given by
\begin{equation}
    \dPS_3(s^+,m_3^2,m_4^2) = \frac{dE_3^* \; dE_7^* \; d\phi_{37}^* \;
		d\phi_+^* \; d\cos\theta_+^*}{256\pi^5} \;,
\end{equation}
with all quantities defined in the $W^+$ rest frame: $E_i^*$ is the energy of
particle $i$, $\phi_{37}^*$ the angle between particles 3 and 7 (the photon),
and $\phi_+^*,\theta_+^*$ the orientation of particle 3 with respect to the
$W^+$.  We now map away the $1/t$, $1/E^*_7$ and $1/(\sqrt{s^+}/2 - E_3^* +
(m_3^2-m_4^2)/2\sqrt{s^+}) \propto 1/(p_4 p_7)$ behaviours of the matrix
element, taking into account the cuts on the photon energy and angle as far as
possible.\footnote{The directions of the momenta are unknown when the energies
are chosen, so we have to err on the safe side; this gives rise to a loss of
efficiency of about 4\% at $\sqrt{s} = 190$ GeV and slightly more at higher
energies.}  Care was taken that the numerical stability is good enough for
electron masses.  The routines implementing these mappings are {\tt mapt},
{\tt mape1} and {\tt mape2}.

\item[{\tt five2a(1),g2a(1)}:] radiation off particle 3 (the outgoing $u$ or
$c$ quark).  These cases are the same as the previous ones with $3
\leftrightarrow 4$.

\item[{\tt five2b(0),g2b(0)}:] radiation off particle 5 (the outgoing $e^-,
\mu^-,\tau^-,d,s$ or $b$).  These routines are obtained from {\tt
five2a(0),g2a(0)} by interchanging the $W^+$ and $W^-$.

\item[{\tt five2b(1),g2b(1)}:] radiation off particle 6 (the outgoing
$\bar{u}$ or $\bar{c}$ anti-quark).  These cases are the same as the previous
ones with $5 \leftrightarrow 6$.

\end{description}

\subsection{Matrix element}

The massless matrix element was taken over from Refs
\cite{Andre&DanielBrems,AndreThesis}.  It uses a massless helicity method,
which breaks down in the collinear region $p_7 \parallel p_{1,2}$.  In this
region we use a collinear approximation described in the next subsection.  The
finite width of the $W$ is introduced by replacing all propagators $1/(p^2 -
m_W^2)$ by $1/(p^2 - m_W^2 + im_W\Gamma_W)$ (without this the cross section is
infinite).\footnote{The effect of using a $p^2$-dependent width is
negligible.}  This is known to break gauge invariance, and is also not a
unique prescription, as it depends on which factors have been cancelled
against the numerator.  The extent of the violation of electromagnetic gauge
invariance had been checked before by changing the photon polarisation vector
to its momentum.  The gauge breaking terms turned out to be small: this test
indeed gives a near-zero amplitude.

We have checked this matrix element against one generated by the Grace package
\cite{Grace} for all the possible final states.\footnote{Due to the long
evaluation time (more than one second on our workstation) this matrix element
is not suitable for use in a Monte Carlo study, given the 13-dimensional phase
space.}  This matrix element includes masses, and gives us the option to test
the contribution of non-resonant, $ZZ$, $Z\gamma$ and $\gamma\gamma$ graphs.
For massless particles and a zero $W$ width the agreement was very good (5-8
decimal places).  For a non-zero $W$ width the two disagree by a small amount
due to the non-unique way the width is introduced in each matrix element.  In
the non-collinear region the mass effects are also small, of order $m^2/s$ as
expected.  In the Grace matrix element we also studied the effect of the gauge
variant terms by explicitly varying the gauge parameters.  The effect is
significantly smaller than the accuracy of our calculation.

The following tests were all performed at an energy $\sqrt{s}=190\GeV$.
For the integrated cross section with 0.1 GeV $< E_\gamma<$ 60 GeV and
no cuts on angles, the two methods agree to better than $\sim$0.1\%, for final
states involving $e$, $\mu$, $u$ or $d$;  better than $\sim$0.25\% for
$\tau$'s, and better than $\sim$2\% for $b$ quarks.  This deviation for
$b$ quarks occurs in the collinear region and will be hidden by jet
fragmentation.  With Grace we can also test the effect of the non-resonant
graphs, which increases the above errors to $\lsim$0.5\% (leptonic
without $ZZ$ and $Z\gamma$ graphs), $\lsim$1\% ($u d c s\gamma$) and
$\lsim$3\% ($u d c b \gamma$).  The errors are somewhat larger in
certain areas of phase space, but only when the amplitude is small.  We have
also tested that the exclusion of $ZZ$ and $Z\gamma$ diagrams in the case of,
e.g., $\mu\nu\mu\nu\gamma$ final states, is not a problem, given a small cut
on such events coming observably from resonant $Z$'s and $\gamma$'s.

\subsection{Mass effects}

To introduce the correct collinear logarithms we multiply the massless matrix
ele\-ment by $\prod_i(q_i q_7)/\prod_i(p_i p_7)$, where the product runs over
all charged particles.  This, however, fails to take into account the double
pole terms, which have a numerator proportional to $m_i^2$ but after
integration give a finite contribution; these are added by hand.

In the case of radiation off the initial state positron or electron (channels
{\tt wg1a} or {\tt wg1b}) the matrix element is replaced in the collinear
region $(p_1 k) < m_e^2$ by the approximation \cite{KleissMass} (in {\tt
getmm1})
\begin{equation}
    |{\cal M}(s)|^2 \to 4\pi\alpha_0 \left( \frac{1}{\xi}\frac{1+\xi^2}{1-\xi}
        \frac{1}{(p_1 k)} - \frac{m_e^2}{(p_1 k)^2} \right) |{\cal
        M}^{(0)}(\xi s)|^2 \;,
\end{equation}
with $\xi = 1 - E_7/E_1$ the splitting variable and ${\cal M}^{(0)}$ the
matrix element without the extra photon.  The mass term is added to the
original matrix element in a much larger cone.  The result agrees to within
$\frac{1}{2}\%$ with the full massive matrix element \cite{Grace}.

When the photon is predominantly radiated off the final state particle $i$
(channels {\tt wg2a} and {\tt wg2b}) we add to the matrix element the double
pole terms (with {\tt getmm2})
\begin{equation}
    |{\cal M}(s)|^2 \to |{\cal M}(s)|^2 - 4\pi\alpha_0 q_i^2
        \frac{m_i^2}{(p_1 k)^2} |{\cal M}^{(0)}(s)|^2 \;,
\end{equation}
where $q_i$ is the charge of particle $i$ in units of the electron charge.
These terms are negligible outside the collinear region, and the sum agrees
excellently (to $\order(m_i^2/s)$) with the full massive matrix element, as
can be seen from the tests cited in the previous section.

\subsection{QCD effects}

Leading order QCD effects have been incorporated in the rate by multiplying
each hadronic $W$ vertex by a factor $(1+\alpha_s(m_W^2)/\pi)$ with
$\alpha_s(m_W^2) = 0.133$.


\section{Results}

\begin{figure}[p]
\unitlength 1bp
\begin{picture}(455,527)(0,-20)
\put(0,0){\strut\epsffile{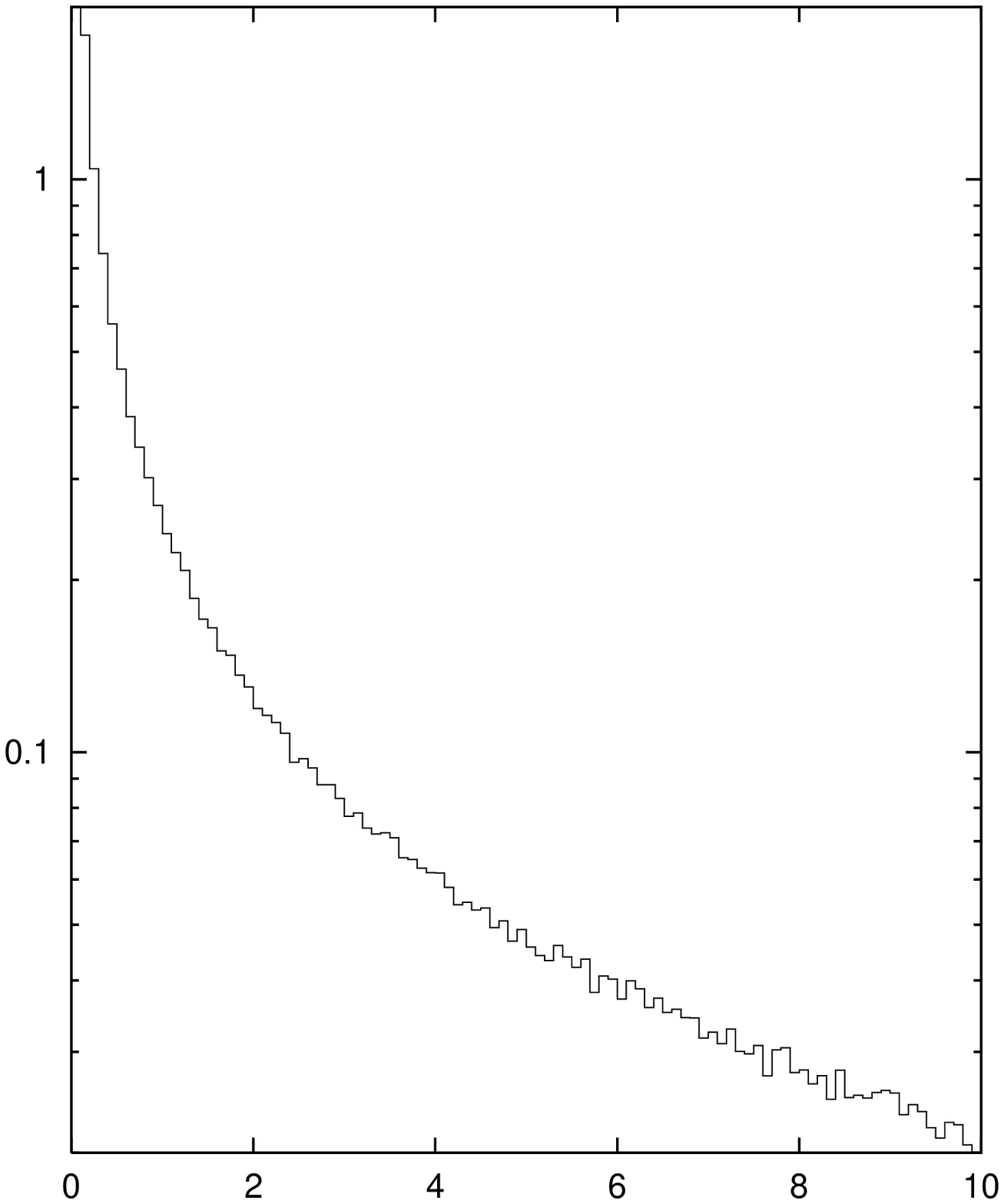}}
\put(170,201){\strut\epsffile{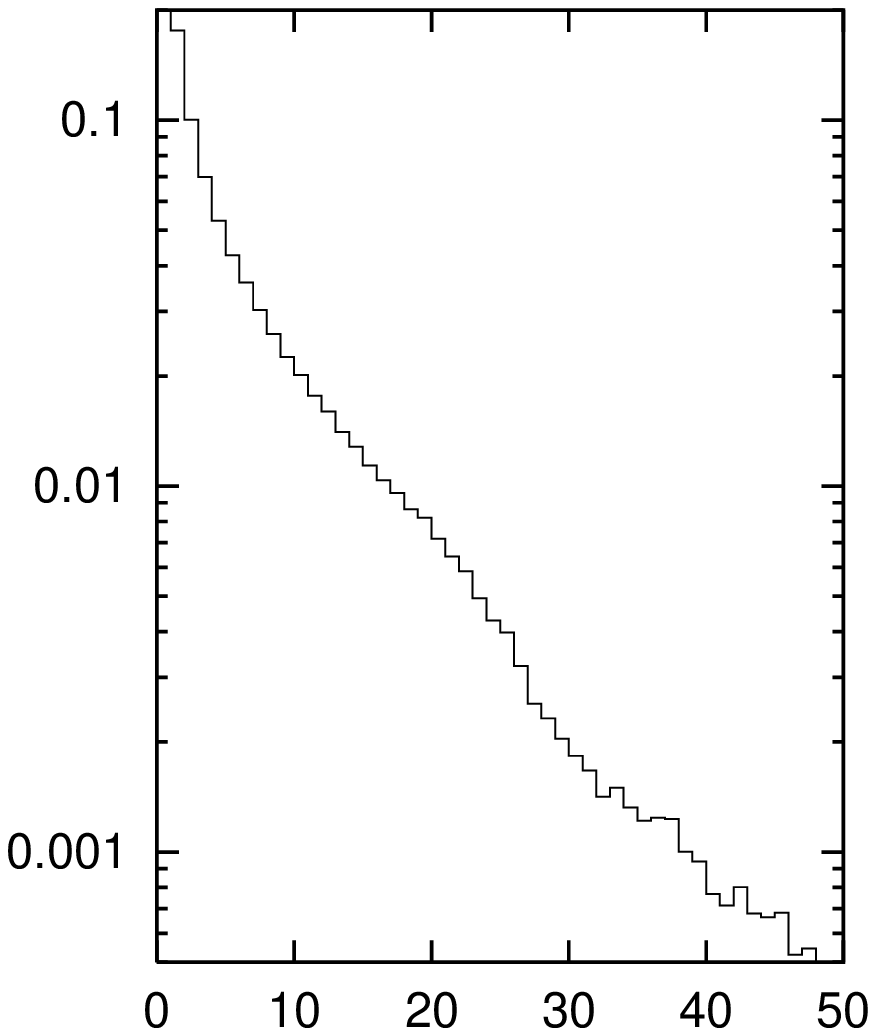}}
\put( 62,497){\makebox(0,0)[tr]{\shortstack{\Large
$\displaystyle \frac{d\sigma}{dE_\gamma}$\\\large\strut$[\PB/\GEV]$}}}
\put(445,0){\makebox(0,0)[tr]{\Large$E_\gamma$ \large$[\GEV]$}}
\end{picture}
\caption[]{The photon energy spectrum at $\sqrt{s} = 190 \GeV$ in the leptonic
channel.}
\label{fig:egam}
\end{figure}

We used the Vegas program to generate the following plots.  The parameters
used are $m_W = 80 \GeV$ and $\Gamma_W = 1.956 \GeV$; we work in the $\alpha$
scheme.  No angular cuts have been applied and collinear initial state
radiation is included.  The photon energy spectrum in the leptonic channel at
$\sqrt{s} = 190 \GeV$ is shown in Fig.\ \ref{fig:egam}.  One would expect some
structure around $E_\gamma \approx \Gamma_W$ as at this energy the $W$ width
starts to function as an infra-red cut-off; however, this is totally washed
out by the radiation collinear to the light external charged particles.

\begin{figure}[p]
\unitlength 1bp
\begin{picture}(455,527)(0,-20)
\put(0,0){\strut\epsffile{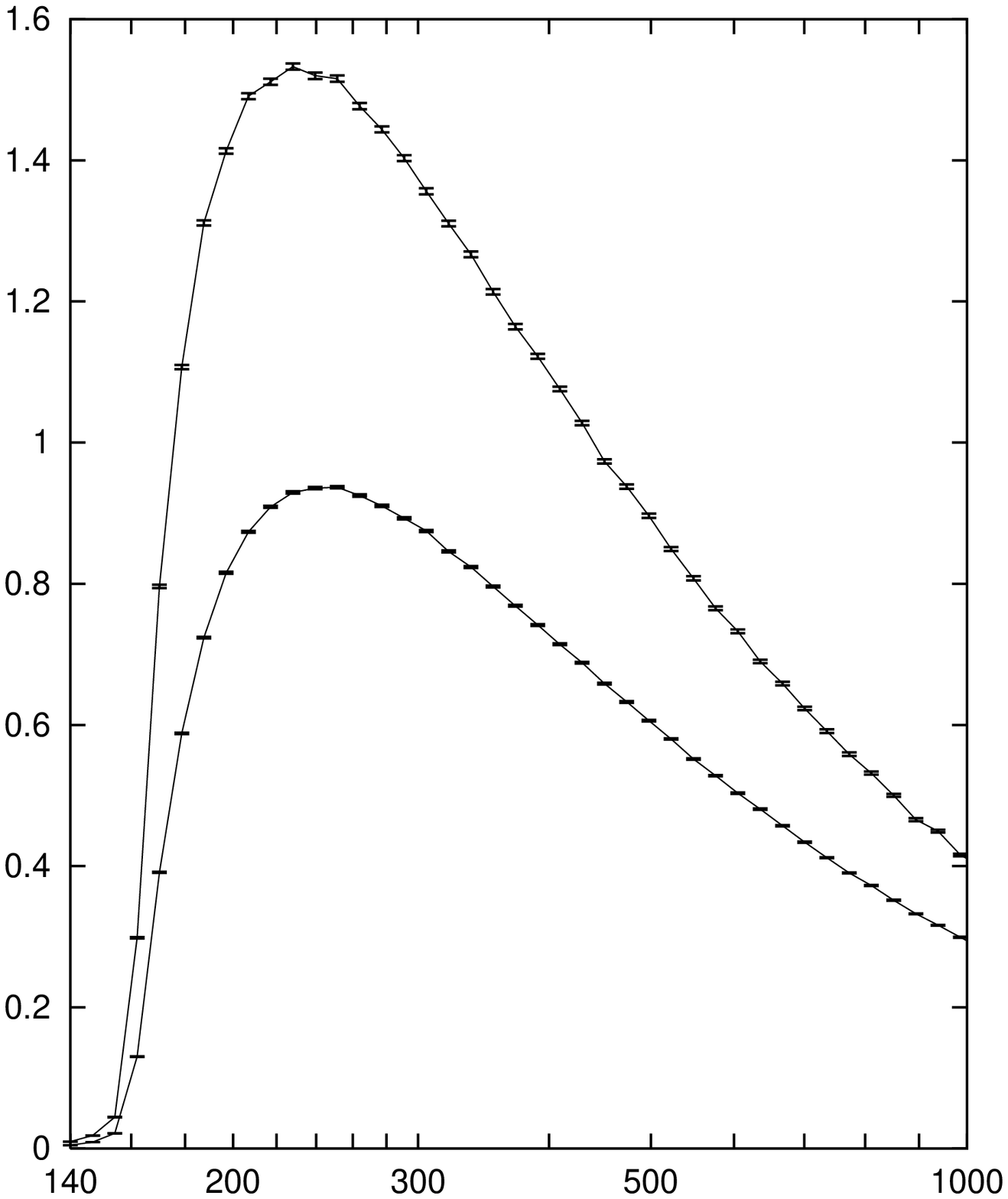}}
\put(260,370){\makebox(0,0)[bl]{\large$E_\gamma > 0.1\GeV$}}
\put(300,200){\makebox(0,0)[tr]{\large$E_\gamma > 1\GeV$}}
\put( 32,497){\makebox(0,0)[tr]{\shortstack{\Large$\sigma$\\
\large\strut$[\PB]$}}}
\put(445,0){\makebox(0,0)[tr]{\Large$\sqrt{s}$ \large$[\GEV]$}}
\end{picture}
\caption[]{The total leptonic cross section for $E_\gamma > 0.1\GeV$ and
$E_\gamma > 1\GeV$.}
\label{fig:cross}
\end{figure}

A plot of the total leptonic cross section with these parameters is given in
Fig.\ \ref{fig:cross}.  To avoid the $Z$ peak we took $E_\gamma < E -
(100\GeV)^2/(4E)$.  In the range of lower photon energy cuts shown, the cross
section scales logarithmically with the cut-off, as it should.  This only
makes sense when this calculation is combined with the virtual and soft
corrections, which will have the opposite logarithmic divergence to give rise
to a finite total cross section.

To give an idea of the attainable accuracy, we computed the cross section at
each energy with 5 iterations of $10^5$ points; this gives a relative error of
0.2\% and takes about 10 minutes per energy.

\section{Usage}
\label{sec:usage}

\subsection{Parameters}
\label{subsec:parameters}

The physics parameters are divided into two classes.  The constants which are
unlikely to change are defined as parameters in the include file {\tt ww.h}.
These include the fermion masses, coupling constants and
CKM matrix elements.
The variable ones have to be defined in {\tt wwfset}: these are the on-shell
$W$ mass (in GeV) {\tt mw} and the physical width of the $W$ {\tt wwidth}.  If
{\tt wwidth} is specified smaller than zero the one-loop on-shell width is
computed.  If {\tt schpar} is zero the width is taken constant, otherwise an
$s$-dependent expression is used.  Finally one has the choice of working in
the $\alpha$-scheme ({\tt schpar}=1) or $G_f$-scheme (2). This does not
influence the final state photon couplings; this coupling constant is always
taken to be $\alpha(0)$.

The next group of parameters concern the decay channels which one wishes to
be included.  The parameter {\tt dectwo} should be defined to be the sum of
{\tt leptonic}, {\tt semileptonic} and/or {\tt hadronic}.  The leptonic decays
of the $W^+$ and $W^-$ can further be specified with {\tt declpp} and {\tt
declpm} respectively, where the possibilities are {\tt electron}, {\tt muon}
and/or {\tt tau}.  There is no possibility to specify the hadronic decays.

Finally one should specify whether one wants additional collinear initial
state photons included by setting {\tt inirad}.

\subsection{Integration}

As the matrix element is infra-red divergent one has to specify (in {\tt
wwfset}) a lower limit on the photon energy of the generated events, {\tt
ecut}.  Another problem occurs at the upper end of the photon energy range: if
the energy of the photons (possibly including collinear initial state photons)
is too large one reaches the $Z$ peak.  As the phase space routines do not
take this peak into account the efficiency of the Monte Carlo will drop
dramatically if this region is included.  The contribution of the $Z$ peak is
negligible ($\order(10^{-4})$, \cite{Andre&DanielBrems}), so this will not
influence the final results.  If {\tt eupcut} is not defined in {\tt wwfset}
it is taken to be 60 GeV, which is reasonable for LEP II.  A warning is
printed if the $Z$ peak is reachable.

Often one is not interested in photons emitted collinear to final state
particles or the beam pipe.  To increase the generation efficiency one can
implement these cuts in the phase space routines by setting {\tt acut} (the
angle between the photon and all charged particles) and {\tt bacut} (the angle
to the beam pipe) to a nonzero number of degrees.  Note that the cone excluded
around the beam pipe is the maximum of these two cuts.

\subsection{Events}
\label{subsec:jetset}

Whenever an event is accepted the routine {\tt spevnt} is called.  This stores
the final state particles in {\tt jetset} structures \cite{jetset73}.  This is
also the place to include a further analysis or detector simulation.  The
four-vectors are stored in {\tt p(0:3,7)}, the dot products in {\tt piDpj(7,7)}
(massive) and {\tt qiDqj(7,7)} (massless) and the particle id's in {\tt
iparti(7)} in jetset codes.

\subsection{Summary of approximations}

As discussed before we made the following approximations in the event
generator:
\begin{itemize}
\item Only $WW$ diagrams are included, no $ZZ$, $\gamma Z$ or $\gamma\gamma$
diagrams which could lead to some of the final states.  With a small cut this
is not a problem.
\item Only the resonant $WW$ diagrams are included; the effect of the
non-resonant diagrams is minute in the case without extra photons
\cite{Andre&DanielBrems,OGamma} and we have tested it to be small here as well.
\item We use a massless matrix element, which has been corrected for collinear
mass effects (the logarithmic and finite double pole
terms).  The neglected $\order(m_f^2/s)$ effects will be unobservable.
\item The $W$ width is introduced in a gauge variant way; however, we have
seen that the numerical effect of this is small in the LEP II region.
\item There should be an upper cut on the photon energy to avoid the $Z$ peak;
the numerical influence of this peak is completely negligible
\cite{Andre&DanielBrems}.
\end{itemize}


\paragraph{Acknowledgements}

We would like to thank Andre Aeppli for making available the code for his
original project.  
We also very much appreciate the use of the Grace package developed by the KEK
theory group.




\appendix
\section{Installation}

The generators have been written in reasonably standard fortran and have been
tested on Sun (both SunOS 4 \& 5), HP 700, DEC {\Large$\alpha$} and NeXT.  The
source code is available with anonymous ftp or WWW from {\tt pss058.psi.ch} in
{\tt /pub/wwf}.  The included makefile should be easily adaptable to your
site.  It defines the targets {\tt wwfmc}: integration with Vegas; {\tt
wwfsimple}: event generation with the simple algorithm; {\tt wwfbases}:
integration with Bases and {\tt wwfspring}: event generation with Spring.
Note that this distribution does not include a copy of Bases/Spring; these can
be obtained from {\tt kekux.kek.jp} in {\tt kek/minami/bases50}.

Some points to note are:
\begin{itemize}
\item On OSF, the command {\tt ranlib} does not exist; please delete these
lines from the {\tt Makefiles}.  (HP-UX gives a laconic message.)
\item HP-UX does not have a {\tt flush} routine used by the version of Vegas
included; edit the {\tt lib/Makefile} to include the replacement {\tt
flush.c}.
\item As usual, the optimisation option on some of the compilers was found to
be less than trustworthy.
\end{itemize}

\end{document}